\documentclass[prb,twocolumn,showpacs]{revtex4}
\usepackage{graphicx,epsfig,color}

\begin{document}

\title{Modification of anatase TiO$_2$(001) surface electronic structure by Au impurity}

\author{E.~Mete$\,^{a,}$\footnote{Corresponding
author: \indent e-mail: emete@balikesir.edu.tr},
O. G\"{u}lseren$\,^b$, and
\c{S}.~Ellialt{\i}o\u{g}lu$\,^c$}

\affiliation{$^a$Department of Physics, Bal{\i}kesir University,
\c{C}a\u{g}{\i}\c{s} Campus, Bal{\i}kesir 10145, Turkey \\
$^b$Department of Physics, Bilkent University, Ankara
06800, Turkey \\
$^c$Department of Physics, Middle East Technical University,
Ankara 06531, Turkey}

\date{\today}

\begin{abstract}
We have used density functional theory calculations based on the
projector augmented wave method to investigate the electronic
structure of Au-incorporated anatase TiO$_2$(001) surface. Due to
the coordination with several level oxygens, Au atoms can be
encapsulated inside TiO$_2$ slab. Au is adsorbed over the surface
Ti--O bond, so called the bridge site on anatase TiO$_2$(001)--1$\times$1
surface. However, for 0.25 ML coverage, Au atoms energetically
prefer to stay at 0.64 {\AA} above the midpoint of the two surface
oxygens which is significantly closer to the surface layer. When
implanted inside the slab for full coverage, Au forms parallel
metallic wires inside TiO$_2$ lattice where interlayer distances
increase due to local segregation. Au brings half-filled
impurity states into the band gap leading to metallization, in
addition to other filled surface and impurity bands within the gap.
These Au-driven Fermi-level-pinning gap states are close to, or even
in some cases inside, the conduction band of the host slab. On the
other hand, if Au is substituted for the surface Ti atom, Fermi
level falls lower in the gap closer to the valence band top.
\end{abstract}


\maketitle

\section{Introduction}

Lately, titania (TiO$_2$) has received an increased attention
since it is considered to be promising for cost-effective
photovoltaic applications. However, high reactivity of TiO$_2$
under only UV light bears a great disadvantage which reduces the
quantum efficiency so that pure TiO$_2$ is not alone sufficient
for a practical system application.~\cite{kim,kowalska,ko,kitano,wang}

McFarland and Tang~\cite{McFarland}, in a recent work, proposed an
Au/TiO$_2$/Ti multilayer photovoltaic device on which photon
absorption occurs in the deposited dye molecules while electron--hole
pairs are created upon light illumination inside the semiconductor
for the conventional solid-state solar cells. This ultrathin
metal--semiconductor junction Schottky diode has driven a particular
attention to gold--titania interface~\cite{ossicini}.

Gold has been proposed to enhance the catalytic activity of anatase
TiO$_2$.~\cite{chen,jung} Au incorporated anatase-based nanocatalysts
have been synthesized for device applications.~\cite{yan,li,grirane}
Moreover, diffusion of gold into anatase polymorph of titania has also
been reported.~\cite{perkas}

The anatase phase of titanium dioxide, although being less
stable than the rutile polymorph, is catalytically more
active~\cite{hengerer,bouzoubaa,thomas}. In this
sense, the surface properties are of major importance. There
are several studies on the topological and electronic structure
of the single crystals of anatase TiO$_2$(001)
surface~\cite{munnix,beltran,lazzeri,calatayud}.
In this paper, we investigated the reconstructive effect of various types of Au
incorporation in the anatase substrate and the role of such an impurity on the
electronic structure of TiO$_2$(001). We considered Au in and on the surface at
quarter and full monolayer (ML) concentrations including Au(Ti) substitutional
cases as well as gold dimer adsorption on the surface. We have discussed their
thermodynamic stabilities for various experimental environments.

\section{Method}

The total energy density functional theory (DFT) calculations have
been performed within the generalized gradient approximation (GGA)
for the exchange--correlation effects via Perdew--Burke--Ernzerhof
(PBE96) functional~\cite{pbe}, using plane-wave basis sets and
the projector augmented waves (PAW) method~\cite{paw1,paw2} as
implemented in the Vienna Ab-initio Simulation Package
(VASP)~\cite{vasp}.

Anatase TiO$_2$(001) stoichiometric surface has been modeled as
a symmetrical slab which constitutes 6 TiO$_2$ layers in a
supercell with a vacuum region of at least 13 {\AA}. A TiO$_2$
layer consists of 3 atomic planes in which bridging oxygen atoms
are out of the level Ti-plane. Our choice regarding the number
of layers represent one of the thickest slab models among
the other theoretical studies~\cite{munnix,beltran,lazzeri,calatayud}.
Moreover, for Au implantation deeper than the fourth atomic plane
we used 8-TiO$_2$-layer slab model to avoid any interaction
between the Au impurities within the same supercell which are
implanted from the bottom and from the top surfaces.

For the gold incorporated surface models we kept ions,
only in the central TiO$_2$ layer, fixed to their bulk positions
because the forces on these remain to be insignificantly small in
all of the calculations. We employed a conjugate-gradients algorithm
to compute the electronic ground state based on the minimization of
the total energy subject to a convergence tolerance of 0.1 meV.
Geometry optimization has been performed, subsequently, by the
reduction of the quantum force on each of the unconstrained ions to
less than 10 meV/{\AA}.

Single particle wavefunctions have been expanded in terms of
plane waves up to a cut-off energy value of 400 eV. The
Brillouin zone (BZ) integrations have been carried out over 32
and 5 special $k$-points sampled in the irreducible wedge for
1$\times$1 and 2$\times$2 surfaces, respectively. We did not impose
any symmetry in our surface calculations.

In thermodynamic equilibrium, the most stable surface composition,
that is entirely surrounded by a thermal bath at temperature $T$
under a given pressure $p$, minimizes the surface Gibbs free energy
$\gamma(T,p)$ (GFE).~\cite{qian,reuter,meyer,timon} Therefore,
relative stability of the phases can be expressed in terms of the
difference $\Delta\gamma(T,p)$ in the GFE of impure surface
structures and relaxed clean surface, as,
\begin{eqnarray*}
\Delta\gamma (T,p)=\hspace{-4mm} && \frac{1}{A} \Big\{
G_{{\rm Au/TiO}_2}(T,p,\Delta n_{\rm Ti},\Delta n_{\rm Au})-
G_{{\rm TiO}_2}(T,p) \\
&& \;\;\;\;\;\; - \Delta n_{\rm Ti}\mu_{\rm Ti}(T,p) -
\Delta n_{\rm Au}\mu_{\rm Au}(T,p)\Big\} \, ,
\end{eqnarray*}
where $G_{{\rm Au/TiO}_2}$ and $G_{{\rm TiO}_2}$ are the GFEs of
the Au incorporated and reference clean TiO$_2$ surface slabs,
respectively. $A$ is the corresponding unit cell area, also
functioning as normalization for different stoichiometries. As the
name suggests, $\Delta n$ is the difference in the number of atoms
from that of the reference surface while $\mu$ denotes the chemical
potential for the referring species. A negative $\Delta\gamma$ value
indicates a more stable structure relative to the clean TiO$_2$
surface. Positive values, on the other hand, give the relative
formation energy that is required to assemble the corresponding
composition.

GFE, in general, is defined by, $G=U+pV-TS$. One can omit $pV$ term
in comparison to the surface energy for the pressures, $p$, under
consideration as well as the $TS$ term since the contribution from
the entropy, $S$, is negligible. Therefore, it can be approximately
expressed only by the internal energy $U$. If one also neglects the
ionic vibrational contributions, $U$ will be equal to the total
energy $E_{\rm total}(N,V)$ obtained by the DFT slab calculations.

Thermodynamic equilibrium is established when the chemical potential
of a given atomic species becomes equal in all phases that come into
contact with each other. In particular, we assume that the surface
structures are in equilibrium with the anatase TiO$_2$ bulk so that
$\mu_{\rm Ti}+2\mu_{\rm O}=\mu_{{\rm TiO}_2}$. Moreover, the
reference energies at the most stable elemental phases of Ti and Au
set the upper bounds for their corresponding chemical potentials.
For instance, the maximum value of $\mu_{\rm Ti}$ can be accessed
in the $hcp$ Ti bulk solid phase. Otherwise, the bulk phase would
be unstable with respect to precipitation of bulk Ti. Similarly,
$\mu_{Au}$ can not be above the chemical potential of $ccp$ gold
bulk. Molecular oxygen defines the upper boundary for $\mu_{\rm O}$
so that $\mu_{\rm O}=\frac{1}{2}E_{{\rm O}_2}$ which corresponds
to an O--rich experimental environment, therefore, defining the
minimum value of $\mu_{\rm Ti}$ through the thermodynamic equilibrium
condition, $\mu_{\rm Ti}+2\mu_{\rm O}=\mu_{{\rm TiO}_2}$.

\section{Results and Discussion}

We systematically studied Au impurities on and inside the
anatase TiO$_2$(001) stoichiometric surface. The role of such
an incorporation on the electronic properties of anatase surface
has been investigated by examining adsorptional, substitutional,
and interstitial impurities at quarter and full coverages.
For these concentrations, we considered gold ions in the
subsurface at the interstitial cavities up to a depth of the
fifth atomic plane, and also substituted them for Ti ions up
to the third Ti plane.

\begin{figure}[b]
\epsfig{file=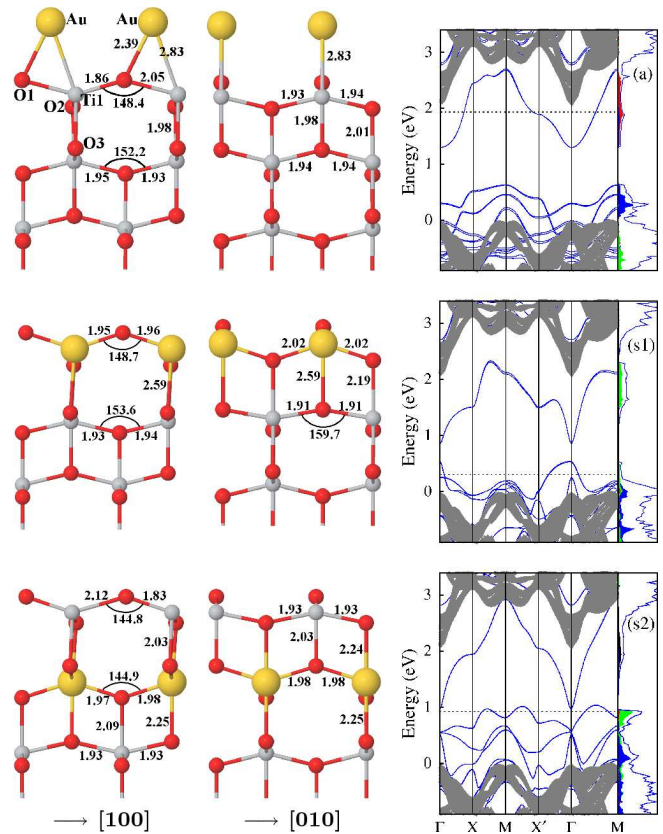}
\caption{(color online) The geometric and electronic structures
of a--$1\times1$ (first row), s1--$1\times1$ (second row), and
s2--$1\times1$ (last row), for the Au/TiO$_2$(001)-1$\times$1
systems. Big lightgray (yellow) balls represent gold. Small dark
(red) and bright (gray) balls denote oxygen and titanium atoms,
respectively. Bond angles and bond lengths are given in degrees
and angstroms, respectively. Energy bands and DOS structures are
described in detail in the corresponding subsections.~\label{fig1}}
\end{figure}

The topological structure and the electronic properties of the
bare anatase TiO$_2$(001) has been discussed in detail
elsewhere~\cite{mete}. Fig.~\ref{fig1} shows the optimized
geometries and corresponding electronic structures of Au--TiO$_2$
combined systems for 1$\times$1 surface. The first two panels to
the left show the side views of the geometric structures along [100]
and [010] directions. In addition, relevant electronic structures
are presented next to these geometries at each row representing a
distinct impurity case. Each ion is labeled with respect to the
atomic plane, that it belongs to, for each of species separately.
For instance, an oxygen at the second oxygen atomic layer is
labeled as O2. Furthermore, bulk termination gives rise to
undercoordinated ions by breaking the axial bonds over the surface.
As a result, O1 and Ti1 become twofold and fivefold coordinated,
thus, also referred as O2c and Ti5c, respectively.

\begin{table*}[t]
\caption{Calculated electronic and structural parameters for
Au--TiO$_2$(001) anatase system: work function, position of
the Fermi energy relative to bulk valence band maximum, Au-depth
relative to surface oxygens, Au--O and Au--Ti distances
for each model labeled in compliance with
Fig.~\ref{fig1}, Fig.~\ref{fig2}, and Fig.~\ref{fig3}.
\label{table1}}
\begin{ruledtabular}
\begin{tabular}{rcccll}
Model & $W$(eV) & $E_{\rm F}$(eV) & $h_{\rm Au}${\AA} & $d_{\rm Au-O}${\AA} & $d_{\rm Au-Ti}${\AA} \\ \hline
a--1$\times$1  & 5.45 & 1.93 & ~~2.14  & 2.39(O1)                               & 2.83(Ti1) \\
s1--1$\times$1 & 6.37 & 0.31 & $-0.51$ & 1.95(O1), 2.03(O2)                     & 3.51(Ti2) \\
s2--1$\times$1 & 6.06 & 0.93 & $-3.22$ & 2.24(O2), 1.98(O3, O4), 2.25(O5)       & 3.23(Ti1), 3.28(Ti3) \\%[2mm]
a--2$\times$2  & 5.02 & 2.02 & ~~1.94  & 2.20(O1)                               & 2.68(Ti1) \\
b--2$\times$2  & 4.29 & 2.60 & ~~0.64  & 2.06(O1), 2.73(O2)                     & 3.06(Ti1) \\
c--2$\times$2  & 4.33 & 2.59 & $-2.71$ & 3.17(O2), 2.09(O3), 2.50(O4)           & 2.85(Ti1), 2.77(Ti2) \\
d--2$\times$2  & 4.49 & 2.57 & $-3.40$ & 2.46(O3), 2.05(O4), 3.13(O5), 2.74(O6) & 2.80(Ti2), 2.94(Ti3)\\
e--2$\times$2  & 4.47 & 2.58 & $-5.24$ & 3.11(O4), 2.06(O5), 2.49(O6), 3.55(O7) & 2.93(Ti2), 2.78(Ti3) \\
s1--2$\times$2 & 6.80 & 0.31 & $-0.27$ & 2.01(O1), 2.03(O2)                     & 3.37(Ti2) \\
s2--2$\times$2 & 6.66 & 0.58 & $-3.18$ & 2.07(O2), 2.59(O3), 2.02(O4), 2.03(O5) & 3.45(Ti1), 3.18(Ti3) \\
aa--2$\times$2 & 6.28 & 1.04 & ~~1.57  & 2.32(Au1--O1), 2.07(Au2--O1)            & 2.66(Au1--Ti1),  2.67(Au2--Ti1) \\
\end{tabular}
\end{ruledtabular}
\end{table*}

\subsection{Au/TiO$_2$(001)-(1$\times$1)}

\subsubsection{Adsorptional case: a--1$\times$1} %{a-1x1 geometry}
Au-adsorptional case, that the first row of Fig.~\ref{fig1}
refers to, has been obtained by relaxing all possible initial
configurations in which Au comes into contact with the active
sites on the support surface. Truly, in this resultant geometry,
Au relaxes to a bridge position on O1 and Ti1 that reflects the
maximum interaction between the Au adsorbate and the surface.
At this minimum energy position O1--Au--Ti1 angle becomes
45.4$^\circ$ with Au--O1 and Au--Ti1 bond distances being 2.39 and
2.83 {\AA}, respectively. The reconstructive effect of the
adsorbate on the lattice remains negligibly small at the
subsurface layers. Au adsorption at 1ML coverage adapts an
ideal-like TiO$_2$ atomic arrangement by saturating the
dangling bonds. Yet, the symmetry breaking in the Ti1--O1
bonds which stems from the relaxation of the clean surface,
is still not lifted as bond lengths read 2.05 and 1.86 {\AA}.
The structural parameters as well as some other physical entities
related to electronic properties such as the surface work function
and Fermi energy relative to bulk valence band maximum (VBM) are
presented in Table~\ref{table1}.

%\subsubsection{a-1x1 band structure}
Au adsorbate brings a two-dimensional impurity state in the gap
closer to the conduction band (CB). Fermi level passes through the
saddle point at X$'$ of this band corresponding to a logarithmic
singularity in the site-projected (local) density of states (LDOS).
This half-filled Fermi-level-pinning state, therefore, leads to
metallization. The reactivity of the O1 site is higher than that of
Ti1, which is also apparent from the LDOS presented on the rightmost
panel of the corresponding electronic band structure in Fig~\ref{fig1}.
A group of four occupied defect states lie distinctly below the Fermi
energy derived from the valence band (VB) possessing dominant O1 character.
The contribution of Ti1 to the total DOS around the Fermi energy is
very weak.

\subsubsection{Substitutional case: s1--1$\times$1} %{s1-1x1 geometry}
The most noticeable effect of Au substitution for surface Ti
(Ti1) at 1$\times$1 structure, also referred as s1--1$\times$1,
is that the distance between the top and the second TiO$_2$ layer
increases substantially. For instance, the Ti1--O3 axial bond
length of 1.96 {\AA} at the clean surface extends to 2.59 {\AA}
for Au--O3. In addition to the increase in the interlayer distance,
this extension also gets the Ti2--O3--Ti2 angle to enlarge from
152.0$^\circ$ to 159.7$^\circ$ at the second TiO$_2$ layer.
Moreover, Au--O1 equatorial bonds along [100] at the surface
layer become almost equal being 1.95 and 1.96 {\AA} as
shown in the second row of Fig.~\ref{fig1}.

%\subsubsection{s1-1x1 band structure}
Unlike to that of the adsorptional case, Au--Ti1 substitution
drives Fermi energy to fall lower in the gap closer to the
valence band top. Because of the partially-occupied nature of
this state which is populated by Au, O1, and chiefly by O2, the
combined system goes into the metallic regime. The upper lying
unoccupied defect state, which takes up a big portion of the gap
to the energies above the Fermi level as a result of the strong
dispersion, shows dominant Au and O2 contribution.

\subsubsection{Substitutional case: s2--1$\times$1} %{s2-1x1 geometry}
The optimized geometry presented in Fig.~\ref{fig1}(s2) for
the second substitutional case (s2--1$\times$1) has been
obtained by relaxing an initial configuration which was formed
by replacing Ti2 with Au in the bare 8-TiO$_2$(001)-layer slab
model. Au impurity sits below the surface in the second TiO$_2$
layer having extended interlayer distances, similar to
s1--1$\times$1 case, of 2.24 and 2.25 {\AA} with the top and
the third layers, respectively. On the other hand, Ti1--O1 bond
lengths remain to be nonequivalent with values of 2.12 and
1.83 {\AA} making a Ti1--O1--Ti1 angle of 144.8$^\circ$.

%\subsubsection{s2-1x1 band structure}
Subsurface substitutional gold sets the Fermi level at 0.93 eV
above the valence band maximum (VBM) pinned by a strongly dispersed
empty defect state, of $\sim$2 eV wide, which descends from the CB.
The second band below, with a width of about 0.5 eV is partially filled
causing metallization of the system. This band is degenerate with a third band
of similar width at $\Gamma$ point, and all these three states are
symmetrical with respect to M point along $\overline{\Gamma{\rm XMX'}\Gamma}$.
The unoccupied first band is originating mainly from the Au--O4 and Au--Au interactions. The
half-occupied second one shows dominant O2 and relatively weaker O4
character. Third band has an LDOS peak stemming from O3. The fourth
band in the gap which crosses the third along most of $\overline{{\rm XMX}'}$
is due to the clean surface (see Fig. 2(a) in Ref. [18]).
The LDOS for the surface oxygen (O1) disperses over VB
similar to that of s1--1$\times$1 implying a charge transfer to
subsurface cation sites. This results in a work function of 6.06 eV.
We have also considered Au substitution for Ti3 (not shown).
The band structure for s3--1$\times$1 has common characteristics
with that of s2--1$\times$1. The first and second bands are very similar,
third is no longer degenerate with the second at $\Gamma$, and
fourth, the surface state, is pushed down towards the VB, and is also
symmetric with respect to M. In quest of finding a trend we checked the band structures
for Au substituted even deeper, namely for Ti5 (s5--1$\times$1) and also considered a
case for bulk substitution. Comparison of these series of cases, Au for Ti1 to bulk Ti,
band structures projected to the same surface Brillouin zone shows that the second
and third bands continue to repel each other, while $E_F$ passes through the second
defect band as before.

\subsubsection{Interstitials at 1$\times$1}
For the interstitial case at 1$\times$1 surface, strong
metal--metal interaction distorts the local lattice structure
drastically due to shortened Au--Au distance of 3.76 {\AA}.
This causes a local segregation that leads to the formation of
parallel metallic Au wires along [010] inside TiO$_2$ lattice
where interlayer distances further increase (from $\sim$2.02 to
3.89 \AA).

\begin{figure*}[htb]
\epsfig{file=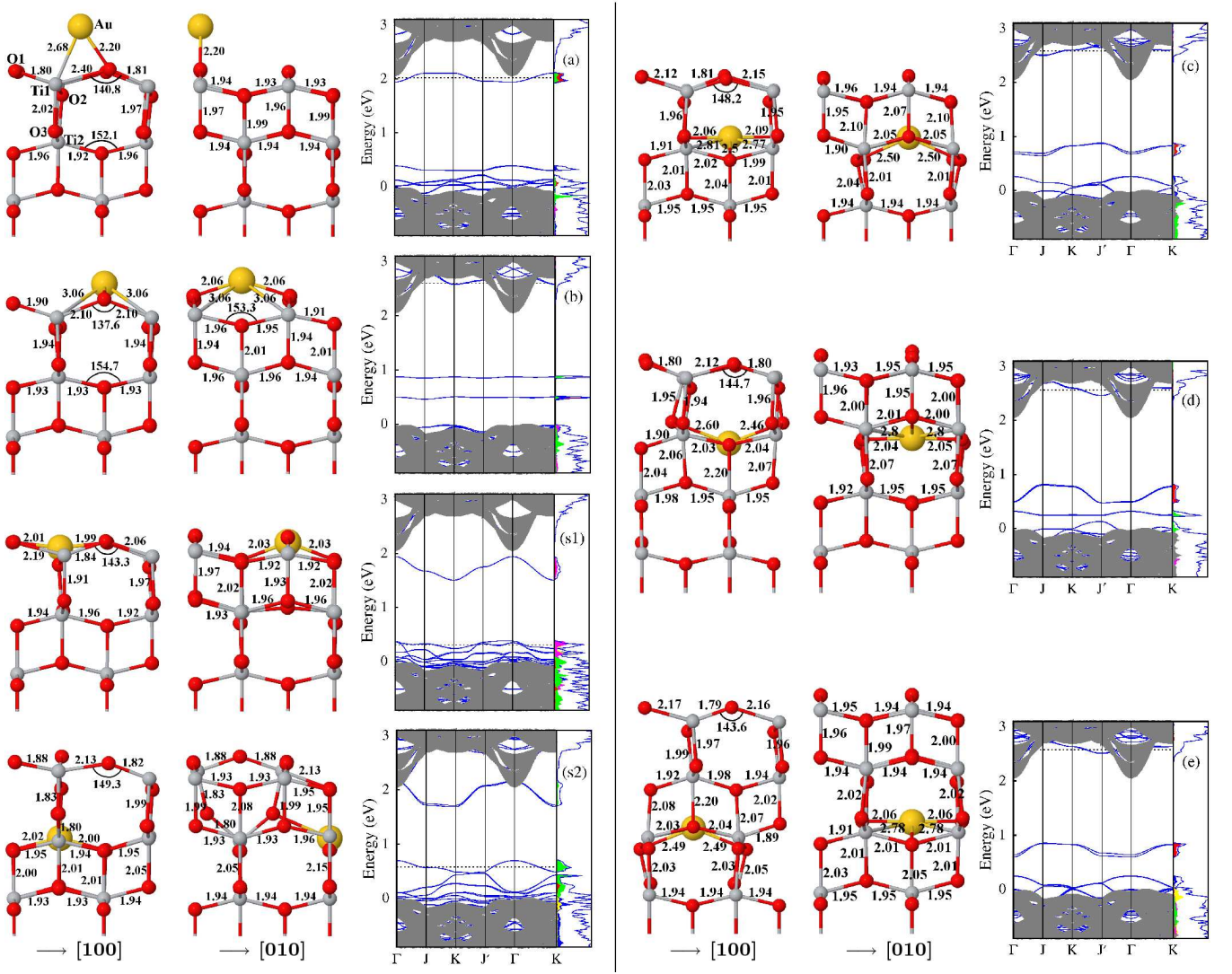}
\caption{Geometric and electronic structures of
Au/TiO$_2$(001)-2$\times$2 systems for the adsorptional,
substitutional, and interstitial gold incorporations.
Label for each case, is given in the DOS panel, should read as,
e.g., a--$2\times2$ for the first row, and so on. Naming convention for the
atoms follow those of Fig.~\ref{fig1}. The bond lengths and angles
are given in angstroms and in degrees, respectively.~\label{fig2}}
\end{figure*}

\subsection{Au/TiO$_2$(001)-(2$\times$2)}

\subsubsection{Adsorptional case: a--2$\times$2} %{a-2x2 geometry}
Similar to the case for the 1ML coverage, single Au adsorbate at
the 2$\times$2 surface is twofold coordinated with O1 and Ti1 at
the bridge position over the Ti1--O1 bond as shown in
Fig.~\ref{fig2}(a). Au promotes the nearest-neighbor Ti1 upward
elongating the Ti1--O3 bond from a clean surface value of 1.96
to 2.02 {\AA}. It also interacts strongly with the nearest O1 which
is elevated up leading to a significant increase in the Ti1--O1
bondlength, from 2.16 to 2.40 {\AA}, over which Au stays at
the minimum energy position. As a result, O1--Ti1--O1 angle gets
wider by 4$^\circ$ compared to the value at the clean surface.
Quarter ML Au adsorption yields significant differences in the
atomic positions from those of the 1ML coverage. For instance,
Au--Ti1 and Au--O1 bond lengths become shorter (2.68 and 2.20 {\AA})
than the values (2.83 and 2.39 {\AA}) for 1ML coverage, respectively.
This can be explained by the increased charge transfer between
the impurity and the support surface due to much weaker Au--Au
impurity interaction at $1/4$ ML coverage.

%\subsubsection{a-2x2 band structure}
For a--2$\times$2 case, energy gap region is characterized by
weakly dispersing six defect states as shown in Fig.~\ref{fig2}(a).
The one lying just below the CB that is half filled by simple
electron counting drives the combined system into metallic state.
This Au--O1 driven state is flat along $\overline{\rm JK}$ and
almost flat along $\overline{{\rm J'}\Gamma}$ and therefore quasi-one
dimensional in nature as seen also in LDOS. The next defect state
that lies 1.56 eV below the Fermi energy at $\Gamma$ being also
quite flat, especially along $\overline{\rm JKJ'}$, has mainly
O2 character. The remaining four states come up slightly above
the VBM with many crossings originating mainly from Au--O1 and
Au--Ti1 interactions.

\subsubsection{Adsorptional case: b--2$\times$2} %{b-2x2 geometry}
In the second adsorptional model, namely b--2$\times$2, Au relaxes
into a very symmetrical position, 0.64 {\AA} above the midpoint
between two O1 ions. It forms two and four equidistant bonds with
those O1 ions and with Ti1s that read 2.06 {\AA} for the former
and 3.06 {\AA}  for the latter, respectively. This isotropic
coordination with two nearest neighbor O1s and with four surface
`Ti's get Ti1--O3 bonds to be aligned parallel to [001]. Hence, each
of the Ti1--O1 bonds become equal to 2.10 {\AA} in length making and
angle of Ti1--O1--Ti1 angle of 137.6$^\circ$. On the other hand, two
Ti2--O2 bonds that are coplanar with Au remain to be slightly skewed
with congruent angles of 3.6$^\circ$ each.

%\subsubsection{b-2x2 band structure}
Fermi level falls into the CB by a resonant defect state causing
metallization for the electronic structure presented in
Fig.~\ref{fig2}(b). Two flat going occupied defect states lie about
0.5 and 0.9 eV above VBM. The higher one of these bands carries
dominant O1 character, whereas the lower one originates from Au--O1
interaction with larger Au mixing, and has some O2 contribution as
well.

\subsubsection{Substitutional case: s1--2$\times$2} %{s1-2x2 geometry}
When substituted for surface titanium as shown in
Fig.~\ref{fig2}(s1), Au relaxes into a position, 0.38 {\AA} above
the original Ti1 lattice point, where it forms two Au--O1 bonds
with slightly unequal lengths of 1.99 and 2.01 {\AA}. It
is also twofold coordinated to two nearest-neighbor O2s at a
distance of 2.03 {\AA} away from each. The inequivalency in the
Ti--O1 bond lengths still remains being 1.84 and 2.06 {\AA} over
the undercoordinated surface oxygen row along [100] next to the
row in which Au substituted. Moreover, Au--O3 separation elongates
to 2.65 {\AA} which can be attributed to Au attaining a nominal
charge state that weakens the charge transfer from Au to O3.
Resulting bond length of Ti2 with this O3 shortens to 1.85 {\AA}
while the other Ti2--O3 distances are all 1.96 {\AA}.

%\subsubsection{s1-2x2 band structure}
A stack of surface bands disperse over an energy range of 0.33 eV
with intercrossings just above VBM. The highest of those is half
occupied setting the Fermi energy at 0.31 eV relative to bulk
valence band top. Therefore, combined system has been predicted
to show metallic behavior. The two sharp LDOS peaks at and just
below the Fermi level in Fig.~\ref{fig2}(s1) are mainly due to O1 which
is bonded to Au. The rather lower part of those surface states,
on the other hand, originate from undercoordinated O1s which are
away from Au site. An empty defect state falls into the gap region
at around 2 eV at $\Gamma$ below CB having a minimum at K. This
band has two saddle points at J and J$'$ which are not at the same
energy because of the unequal Au--O1 bondlengths along [100].
Corresponding LDOS peaks come out as a result of excess anti-bonding
charge localized around gold.

\subsubsection{Substitutional case: s2--2$\times$2} %{s2-2x2 geometry}
Au substitution for Ti2 causes local disturbances on the nearby
O3s as a result of the excess charge embedded by and localized
around the impurity site. Equatorial Au--O3 bonds on (100) plane
extend to 2.59 {\AA} leading to a significant dislocation of O3s
from their lattice positions. This subsurface reconstruction
induces stress on Ti1--O3 interaction that flips up Ti1--O2 bonds.
Then, the displaced O2 levels itself with O1 atomic plane as shown
in Fig.~\ref{fig2}(s2). On the other hand, subtle atomic
rearrangements have been observed upon Au substitution for Ti2 over
the planes perpendicular to [010] direction. In this structure,
Ti1--O1 bonds are still unequal (2.13 and 1.82 {\AA}) making a
Ti1--O1--Ti1 angle of 149.3$^\circ$. Au substitution
for a third layer Ti cation, namely s3--2$\times$2, dislocates
the nearby O3s in the same way as s2--2$\times$2 does. This suggests
that Au--Ti replacement for deeper lying cations will cause similar
local reconstruction.

%\subsubsection{s2-2x2 band structure}
The electronic structure of s2--2$\times$2 model in
Fig.~\ref{fig2}(s2) have similar characteristics with that of
s1--2$\times$2. Both of them has an unoccupied defect state arising
chiefly from Au impurity. However, the one for the s2 case lies
higher and so partly inside CB at around $\Gamma$. It is slightly
dispersed along $\overline{\rm JK}$ and $\overline{{\rm J}'\Gamma}$
causing the band to be a quasi-1D band as opposed to the quasi-2D
band of s1 case. This also shows itself in their LDOS structures. An O2 driven
half-filled defect state having a maximum at $\Gamma$ has been hived
off from the lower lying group of surface-like bands. Therefore,
Fermi level is at 0.58 eV with respect to bulk valence band top. This
is related to the surface oxygen, O2, that is bonded to Au. The next
band down is also of O2 character, however, it gets additional
contributions from the surface oxygen, O2, raised to the level of
O1s, as well. In comparison with those of s1--2$\times$2, the
stack of surface bands disperse over a relatively higher energy
range of 0.47 eV just above VBM. They are due mainly to axial Au--O2
interaction. LDOS also shows contributions to bands slightly below
VBM primarily from surface oxygens. As in the 1$\times$1 case we have
examined the deeper Au substitutions for up to bulk Ti which lead the
half-filled defect state to move about 1 eV upward closer to the
empty band in the gap.

\subsubsection{Interstitials at 2$\times$2}
Au can be encapsulated at the interstitial cavities in the
subsurface layers starting from the fourth atomic layer for
2$\times$2 surface as opposed to 1$\times$1. The first three
possible structures are presented in Fig.~\ref{fig2}(c)--(e).
As common to all of them, Au relaxes into the midpoint between
the two level oxygens where it also establishes equidistant
coordination with four nearest neighbor `Ti's that lie at the
closest Ti atomic plane. Au interstitial causes local
disturbances such that two second nearest-neighbor oxygens at
the preceding or succeeding O layers get slightly repelled out
from their bulk lattice positions due to the induced stress
incorporated by the excess charge at the impurity site. Au
implantation to the internal cavities can not help lifting the
symmetry breaking in the Ti1--O1 bond distances that exists in
the case of the bare anatase surface.

%\subsubsection{interstitials at 2x2 band structure}
The interstitial Au impurities in TiO$_2$(001) surface, corresponding
to Fig.~\ref{fig2}(c)--(e), show similar electronic characteristics
since they represent equivalent local environments. For all of them,
Fermi level falls in CB due to an impurity driven state that lies in
the energy gap region along $\overline{\rm JKJ'}$ and partly along the
$\overline{{\rm K}\Gamma}$ segment. The fully occupied quasi-1D defect
state just below 1 eV disperses in accordance with the spatial alignment
of Au--O coordination. Therefore, its maximum at ${\rm J'}$ for
c--2$\times$2 and at J for d--2$\times$2 and e--2$\times$2, (doubly)
alternates with Au--O bond orientation. Lower lying nearly flat state
in case--d is due primarily to O2 oxygens at the surface. For the cases
c and especially e, the stack of bands just above VBM are very similar
to those of the clean surface (see Fig. 2(a) in Ref. [18]), because the
region of disturbance due to the interstitial Au is away from the first
TiO$_2$ layer for these cases, and for cases with deeper interstitials.

\section{Bader analysis}
We analyzed the electronic charge density using the atom in
molecule (AIM) theory with a grid-based algorithm~\cite{sanville}.
Computational Bader charge results for near surface Ti and O atoms,
and also for the Au impurity, are presented in Table~\ref{table2}
for the combined systems as well as for the clean surface. For
Au--TiO$_2$(001) structures, atomic charge states are provided for
O and Ti ions that are closest to gold site in order to better
describe the local disturbance of the electronic density, posed by
the impurity.

\begin{table}[b]
\caption{Bader charge analysis of Au--TiO$_2$(001) anatase
systems. Atom labels follow Fig.~\ref{fig1}, Fig.~\ref{fig2},
and Fig.~\ref{fig3}. The values represent the valence charge
states for lattice atoms that are closest to Au impurity site.
\label{table2}}
\begin{ruledtabular}
\begin{tabular}{rcccccc}
Model           & O1 & O2 & O3 & Ti1 & Ti2 & Au \\ \hline
clean           & $-$1.26 & $-$1.34 & $-$1.33 & +2.61 & +2.64 & --- \\
a--1$\times$1   & $-$1.20 & $-$1.34 & $-$1.33 & +2.58 & +2.64 & $-$0.03 \\
s1--1$\times$1  & $-$0.70 & $-$0.85 & $-$1.18 &  ---  & +2.64 & +1.38 \\
s2--1$\times$1  & $-$1.24 & $-$1.17 & $-$0.94 & +2.58 &  ---  & +1.54 \\
a--2$\times$2   & $-$1.15 & $-$1.34 & $-$1.32 & +2.55 & +2.65 & $-$0.04 \\
b--2$\times$2   & $-$1.18 & $-$1.35 & $-$1.33 & +2.57 & +2.62 & +0.34 \\
c--2$\times$2   & $-$1.26 & $-$1.33 & $-$1.24 & +2.55 & +2.56 & +0.34 \\
d--2$\times$2   & $-$1.25 & $-$1.34 & $-$1.30 & +2.60 & +2.56 & +0.34 \\
e--2$\times$2   & $-$1.26 & $-$1.34 & $-$1.34 & +2.60 & +2.58 & +0.38 \\
s1--2$\times$2  & $-$0.95 & $-$1.11 & $-$1.26 &  --- & +2.63 & +1.39 \\
s2--2$\times$2  & $-$1.20 & $-$1.03 & $-$1.27 & +2.63 &  --- & +1.32 \\
aa--2$\times$2 (Au1)  & $-$1.23 & $-$1.35 & $-$1.32 & +2.56 & +2.64 & $-$0.07 \\
aa--2$\times$2 (Au2)  & $-$1.12 & $-$1.35 & $-$1.33 & +2.54 & +2.64 & +0.05 \\
\end{tabular}
\end{ruledtabular}
\end{table}

The non-ionic nature of Ti--O bonding in TiO$_2$ stoichiometry,
due to below-nominal charge states of Ti cations and O anions,
has been previously shown~\cite{mete,calatayud2}. Bulk ions
accumulate $Q_{\rm O}=-1.33e$ and $Q_{\rm Ti}=+2.66e$
Bader charges while these values are slightly lower for the
corresponding surface layer species, being $Q_{\rm O1}=-1.26e$
and $Q_{\rm Ti1}=+2.61e$, due to bulk termination.

Bader charge results presented in Table~\ref{table2} for
the impurity neighboring ions demonstrate the local disturbance
of the electronic density. The fluctuation in the charge density
around the cation sites remains minimal upon Au incorporation.
For interstitial and adsorptional cases, gold seems to weakly
bound to the lattice because of a limited charge transfer from
Au to nearby oxygens. We obtained relatively smaller deviations
in the charge states of deeper lying atoms implying a weaker
contribution to surface electronic properties. When gold is
substituted for Ti cations it induces deviation in the electronic
density also around the second nearest neighbor oxygens owing to
its spatially wider spread wavefunction. For instance, Au--Ti1
substitution in s1--1$\times$1 charges the third layer oxygen
to $Q_{O3}=-1.18e$ that is smaller than its reference value
of $Q_{O3}=-1.33e$ at the clean surface.

In comparison with their reference values at the clean surface,
a lower amount of charge accumulation at the anions around
the impurity site has been obtained for all Au--TiO$_2$(001)
structures in Table~\ref{table2}. This elicits a weak
polarization in the covalency between gold and oxygen at
this surface. The Au--O and Ti--O interaction strengths
can be compared based on the charge transfers in between.
The electron depletion from Ti to O is clearly much larger than
that from Au to O suggesting that Ti--O bond is stronger.

\subsection{Au dimer on 2$\times$2}

\begin{figure}[t] \vspace{4mm}
\epsfig{file=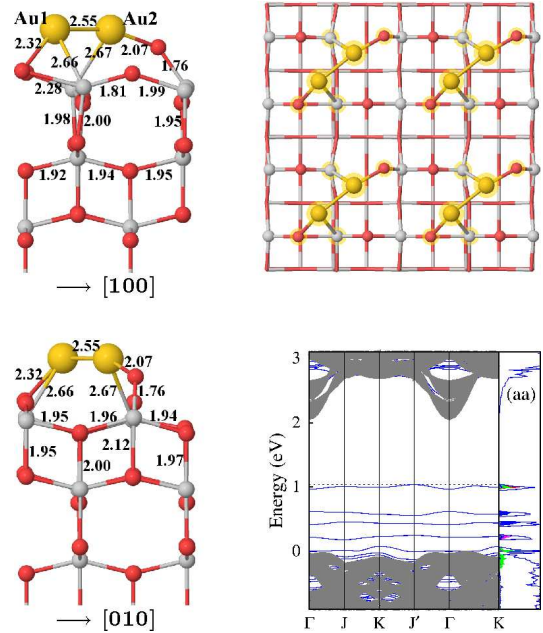,width=7cm} \caption{(color online)
Au dimer on anatase TiO$_2$(001)-2x2 surface (aa-2$\times$2).
Side views of the impurity embedded slab model are shown in the
left panel where all distances are given in {\AA}. Top view of gold
dimers on this surface as well as the corresponding electronic
structure are presented on the right panel.
\label{fig3}}
\end{figure}

The weakness of Au--O bond polarization, in and on TiO$_2$ surface,
compared to that of Ti--O opens the possibility of an important
interaction between two adjacent Au atoms. Clearly, Au driven
impurity bands disperse relatively stronger when they are
incorporated to 1$\times$1 surface unit cell in Fig~\ref{fig1}.
Therefore, we considered all possible adsorption configurations of Au dimer on 2$\times$2
surface in order to investigate the effect of Au--Au interaction
on the electronic structure of the combined system. Fig.~\ref{fig3}
shows the minimum energy Au dimer structure on the anatase
TiO$_2$(001)-2$\times$2 surface. This is obtained from an initial
configuration in which one Au atom is placed over the Ti1--O1 bond
in bridge position and the other one is located over the next
Ti1--O1 bond on the back Ti--O--Ti row. They are attracted to
each other due to Au-Au interaction. Relatively speaking with
respect to the structure shown in left-top of Fig~\ref{fig3}, as
they come closer reducing the dimer length to 2.55 {\AA} the one at
the back row (Au2) pulls the nearest neighbor O1 off its lattice
position up by 0.91 {\AA}. However, the O1 in interaction with the
Au1 at the first row is elevated by only 0.06 {\AA}. Surprisingly,
any geometry that the surface oxygens bonded symmetrically to the
dimer is energetically unfavorable.

We also examined the possibility of a dissociative adsorption for
various configurations which corresponds to 0.5 ML coverage on this
surface. Our results suggest that such a dimer dissociation on the
anatase surface is energetically not preferable signifying the
strength of Au--Au attraction compared to Au--O and Au--Ti
interactions.

Electronically, gold dimer adsorption at 2$\times$2 unit cell drives
the anatase (001) surface into a semiconducting state such that it
reduces the band gap by 1.04 eV relative to anatase bulk VBM. The
impurity band just below the Fermi energy has a minimum at $\Gamma$
and a maximum at J$'$ which gives rise to an indirect gap. This
otherwise almost flat going state lies higher in energy from a
group of five surface-like bands with a separation of 0.34 eV at
$\Gamma$. It is mainly characterized by Au--Au and Au1--O1
interactions with a partial mixing from the unsaturated surface
oxygens. The second defect states disperses conjugately to the
first one having a maximum at $\Gamma$ and a minimum at J$'$.
The surface oxygen that is bonded to Au2 mainly contributes to the
second and the third defect states. The next impurity band that has
a minimum at J and a maximum at J$'$ originates from Au2-O1 bonding.
The LDOS of unsaturated O1 ions extend over the VB, also giving
partial contribution to all of the defect states.

\section{Thermodynamic stability}
We analyzed the thermodynamic stability of gold incorporated
TiO$_2$ systems, considered in this work, assuming Au and Ti
exchange between the substrate and the surrounding gas phase to
account for the formation of substitutional cases.

\begin{figure} \vspace{4mm}
\epsfig{file=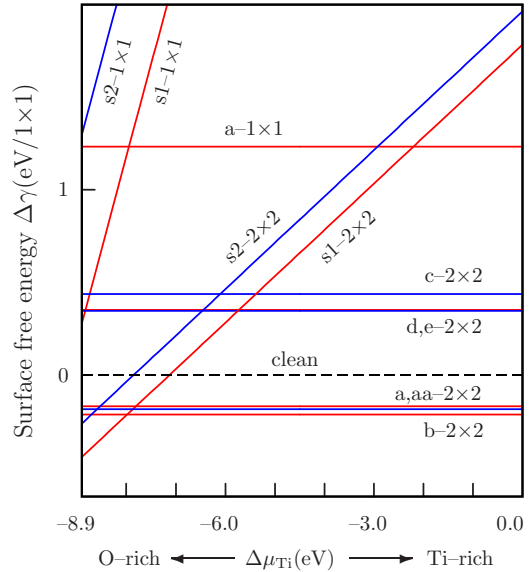,width=7cm} \caption{(color online)
Normalized relative surface Gibbs free energies of
Au--TiO$_2$(001)-1$\times$1 and Au--TiO$_2$(001)-2$\times$2
structures as a function of and over the allowed range of Ti
chemical potential. Au impurity ions are chosen to be in
thermodynamic equilibrium with $ccp$ Au bulk phase.
\label{fig4}}
\end{figure}

We assume experimental conditions so that Au is chosen to be in
equilibrium with its metallic bulk phase and redefine the zero point
of energy at the maximal value of $\mu_{\rm Ti}$ by introducing
$\Delta\mu_{\rm Ti}=\mu_{\rm Ti}-\mu_{\rm Ti}^{\rm bulk}$. The
relative surface GFEs, $\Delta\gamma$, (or the formation energies)
are plotted as a function of $\Delta\mu_{\rm Ti}$ in Fig.~\ref{fig4}
over its allowed range of values for 12 surface structures
considered in this study. The lower limit for Ti chemical
potential has been obtained to be $-8.90$ eV due to the fact
that the surface must keep in thermal equilibrium with the
bulk anatase TiO$_2$ phase.

Our prescription of stability of the phases can predict the
structural reconstructions rather than thermally stimulated
formations due to the approximation made by omitting the
entropy terms in the surface GFE.

The results presented in Fig.~\ref{fig4} suggest that Au
incorporation at 1$\times$1 surface is not favorable. For instance,
Au substitution for the second layer Ti at 1$\times$1 anatase (001)
surface (s2--1$\times$1) is energetically the least preferable phase
among the structures considered. Similarly, s1--1$\times$1 case is
also thermodynamically unstable. Their formation energies further
increase linearly toward Ti--rich experimental conditions. Au
adsorption for 1 ML coverage would require an energy of 1.23
eV/1$\times$1 relative to that of the clean surface.

Au interstitials in the 2$\times$2 surface appears to be unfavorable
compared to bulk terminated bare anatase TiO$_2$(001) by the
formation energies of 0.44, 0.35 and 0.35 eV/1$\times$1 for c, d,
and e cases, respectively. Surprisingly, Au implantation
into the interstitial cavities deeper than the fourth atomic layer
is slightly more preferable at this surface. These calculated formation
energies for Au interstitials in anatase surface are in good agreement
with the experimental results of Perkas~\textit{et al.}~\cite{perkas}.
In an attempt to explain Au induced crystallization (IC) of anatase 
through gold insertion into TiO$_2$ which requires thermal 
treatment at about 80$^\circ$C by sonication, they stated that 
Au diffuses into the support layers and forms and intermixed layer 
through a multiphase diffusion across the gold support interface.

Substitutional gold impurities tend to be linearly more unstable
toward Ti-rich conditions. It is also reasonable to see that Au
substitution for a second Ti layer cation is energetically more
unstable than that for the surface layer Ti over the full range of
allowed $\mu_{\rm Ti}$ values. On the other hand, Au(Ti) substitution
has been expected to be more preferable in a Ti-poor environment.
Indeed, both of the substitutional cases become more stable than the
clean relaxed surface under O-rich conditions when the substrate
establishes thermodynamic equilibrium with the surrounding oxygen gas
phase. Fig.~\ref{fig4} shows that such a surface structure can be
realized under the limits, $\Delta\mu_{\rm Ti}<-7.84$ eV and
$\Delta\mu_{\rm Ti}<-7.14$ eV for s2 and s1 cases, respectively.
Furthermore, Au(Ti) substitution for the surface Ti atom proves
to be the most stable phase for $-8.90<\Delta\mu_{\rm Ti}<-7.98$ eV
that corresponds to a Ti-poor environment.

1ML gold adsorption is not preferable with a formation energy of
1.23 eV relative to that of the clean surface. On the other hand,
Au adsorbates at 2$\times$2 unit cell as well as the dimer structure
in Fig.~\ref{fig3} happen to be even more stable than the formation
of bulk terminated bare (001) surface which is known to be under 
tensile stress~\cite{lazzeri2} whereby Au bonding causes a lowering 
of the surface energy by releasing the surface strain~\cite{Vittadini}
We calculated the corresponding relative surface GFEs as $-0.17$, $-0.21$ 
and $-0.18$ eV/1$\times$1 for a--2$\times$2, b--2$\times$2 and 
aa--2$\times$2 structures, respectively. Therefore, highly symmetrical 
structure in Fig.~\ref{fig2}(b) appears to be the most stable phase over 
a wide range of experimental situations within $-7.84<\Delta\mu_{\rm Ti}<0$
from Ti-low to Ti-rich equilibrium conditions. However, the
differences between the computed GFE values of these three adsorption
modes are, surprisingly, not significantly large and can be accepted
within the limits of a computational accuracy. In this sense, one cannot
conclude that a--2$\times$2 and aa--2$\times$2 are considerably less
stable than b--2$\times$2 is. The results rather suggest that these
three phases might co-exist in an experimental situation where,
interestingly, single Au atom adsorption is metallic while dimer
structure is semiconducting.  Therefore, overall behavior is
expected to be conducting in nature.

\section{Conclusions}
In conclusion, we studied the role of Au mediated lattice
relaxations as well as the effect of the impurity itself on
the electronic structure of anatase TiO$_2$(001) for 1$\times$1
and 2$\times$2 surface unit cells. We, additionally, considered
gold dimer adsorption on TiO$_2$(001)-2$\times$2 to account
for the strength of Au--Au interaction over Au--O attraction.

In comparison with Ti, due to its spatially wider spread
wavefunction, gold maintains a relatively larger distance to the
oxygens with which it interacts. In addition, Au exhibits an ample
coordination with the neighboring oxygens and, therefore, disturbs
local lattice structure. Electronically, an Au adsorbate transfers
a limited amount of charge to the nearest neighbor oxygens that
reduces the ionicity leading to a relatively weaker
impurity--oxygen interaction. On the other hand, Au can attain higher
valence charge states in the case of interstitial and substitutional
incorporations due to increased coordination number with the
neighboring oxygens. For 1ML impurity concentration, gold
interstitial enforced local distortion causes significant
reconstruction in the form of interlayer segregation due to shortened
Au--Au interaction distance. These impurity mediated disturbances
derive new defect states and strongly disperse the existing surface
bands.

Single Au impurity brings a half-filled impurity state into the
band gap of TiO$_2$(001) which pins the Fermi level leading to
metallization, in addition to other filled surface and impurity
driven bands within the gap. In the adsorptional and interstitial
cases, this state is derived from the CB. For Au interstitials as
well as for b--2$\times$2, Fermi energy falls inside the CB. However,
substitutional Au makes it up close to the VB also deriving empty
impurity states higher in energies toward the CB. The dispersion
of the defect states depend on the impurity concentration and the
Au--O interaction strength. Therefore, 1$\times$1 cases exhibit
the strongest dispersions, and the adsorptional Au incorporations
at 2$\times$2 surface lead to flat-like defect states which
represent the weakest Au--O interactions.

All Au--incorporated TiO$_2$(001) structures at 1$\times$1
surface unit cells are unstable. Under strongly oxidizing, Ti--poor
conditions gold tends to substitute surface Ti ions at 1/4 ML
coverage. On the other hand, for a wide range of experimental
situations, from moderate to strongly reducing conditions, Au
adsorption is thermodynamically more stable at the 2$\times$2
support surface. Single Au adsorbates at this coverage bring out
new acceptor sites as they reduce the surface leading to
metallization. Therefore, an additional gold adsorption prefers the
dimer structure rather than being dissociative. Gold dimer supported
by the slab at 2$\times$2 surface is almost as stable as the single
Au adsorption signifying the strength and the role of Au--Au
interaction which drives the system into semiconducting regime.
Consequently, in an experimental situation which realizes the
co-existence of these two  phases, the overall behavior is expected
to be conducting.

\begin{acknowledgments}
EM and {\c{S}}E acknowledge financial support from T\"{U}B\.{I}TAK, The
Scientific and Technological Research Council of Turkey (Grant no:
TBAG 107T560). In conjunction with this project, computational
resources were provided by ULAKB\.{I}M, Turkish Academic Network. \&
Information Center. OG acknowledges the support of Turkish Academy
of Sciences, T\"{U}BA.
\end{acknowledgments}


\begin{thebibliography}{99}
\bibitem{kim} S. Kim, S-J. Hwang, and W. Choi, J. Phys. Chem. B
\textbf{109}, 24260 (2005).

\bibitem{kowalska} E. Kowalska, H. Remita, C. Colbeau-Justin, J. Hupka, and J. Belloni,
J. Phys. Chem. \textbf{112}, 1124 (2008).

\bibitem{ko} K. Ko, Y. Lee, and J. Jung, J. Colloid and Interf. Sci. \textbf{283}, 482 (2005).

\bibitem{kitano} M. Kitano, M. Takeuchi, M. Matsuoka, J.\,M. Thomas, and M. Anpo,
Catal. Today \textbf{120}, 133 (2007).

\bibitem{wang} Y. Wang and D.\,J. Doren, Solid State Commun. \textbf{136}, 186 (2005).

\bibitem{McFarland} E.\,W. McFarland and J. Tang, Nature \textbf{421}, 616 (2003).

\bibitem{ossicini} I. Marri and S. Ossicini, Solid State Commun., \textbf{147}, 205 (2008).

\bibitem{chen} M.\,S. Chen and W. Goodman, Science \textbf{306}, 252 (2004).

\bibitem{jung} J.\,M. Jung, M. Wang, E.\,J. Kim, and S.\,H. Hahn, Vacuum \textbf{82}, 827 (2008).

\bibitem{yan}  W. Yan, B. Chen, S.\,M. Mahurin, V. Schwartz, D.\,R. Mullins,
A.\,R. Lupini, S.\,J. Pennycook, S. Dai, and S.\,H. Overbury,
J. Phys. Chem. B \textbf{109}, 10676 (2005).

\bibitem{li} J. Li and H.\,C. Zeng, Chem. Mater. \textbf{18}, 4270 (2006).

\bibitem{grirane} A. Grirane, A. Coma, H. Garcia, Science \textbf{322}, 1661 (2008).

\bibitem{perkas} N. Perkas, V.\,G. Pol, S.\,V. Pol, and A. Gedanken,
Crystal Growth \& Design, \textbf{6}, 293 (2006).

\bibitem{hengerer} R. Hengerer, B. Bolliger, M. Erbudak, and M.
Gr\"{a}tzel, Surf. Sci. \textbf{460}, 162 (2000).

\bibitem{bouzoubaa} A. Bouzoubaa, A. Markovits, M. Calatayud, and C.
Minot, Surf. Sci. \textbf{583}, 107 (2005).

\bibitem{thomas} A.\,G. Thomas, W.\,R. Flavell, A.\,K. Mallick, A.\,R. Kumarasinghe,
D. Tsoutsou, N. Khan, C. Chatwin, S. Rayner, G.\,C. Smith, R.\,L. Stockbauer,
S. Warren, T.\,K. Johal, S. Patel, D. Holland, A. Taleb, and F. Wiame,
Phys. Rev. B \textbf{75}, 035105 (2007).

\bibitem{munnix} S. Munnix and M. Schmeits, Phys. Rev. B \textbf{30}, 2202 (1984).

\bibitem{beltran} A. Beltr\'{a}n, J.\,R. Sambrano, M. Calatayud, F.\,R. Sensato,
and J. Andr\'{e}s, Surf. Sci. \textbf{490}, 116 (2001).

\bibitem{lazzeri} M. Lazzeri, A. Vittadini, and A. Selloni, Phys. Rev. B \textbf{63}, 155409 (2001).

\bibitem{calatayud} M. Calatayud and C. Minot, Surf. Sci. \textbf{552}, 169 (2004).

\bibitem{pbe} J.\,P. Perdew, K. Burke, and M. Ernzerhof,
Phys. Rev. Lett. \textbf{77}, 3865 (1996).

\bibitem{paw1} P.\,E. Bl\"{o}chl, Phys. Rev. B \textbf{50}, 17953 (1994).

\bibitem{paw2} G. Kresse and D. Joubert, Phys. Rev. B \textbf{59}, 1758 (1999).

\bibitem{vasp} G. Kresse and J. Hafner, Phys. Rev. B \textbf{47}, 558 (1993).

\bibitem{qian} G.-X. Qian, R.\,M. Martin, and D.\,J. Chadi, Phys. Rev. B \textbf{38}, 7649 (1988).

\bibitem{reuter} K. Reuter and M. Scheffler, Phys. Rev. B \textbf{65}, 035406 (2001).

\bibitem{meyer} B. Meyer, Phys. Rev. B \textbf{69}, 045416 (2004).

\bibitem{timon} V. Timon, S. Brand, S.\,J. Clark, M.\,C. Gibson, and R.\,A. Abraham,
Phys. Rev. B \textbf{72}, 035327 (2005).

\bibitem{mete} E. Mete, D. Uner, O. G\"{u}lseren, and \c{S}. Ellialt{\i}o\u{g}lu,
Phys. Rev. B \textbf{79}, 125418 (2009).

\bibitem{sanville} E. Sanville, S.\,D. Kenny, R. Smith, and G. Henkelman,
J. Comp. Chem. \textbf{28}, 899 (2007).

\bibitem{calatayud2} M. Calatayud, P. Mori-S\'{a}nchez, A. Beltr\'{a}n, A. Mart\'{i}n Pend\'{a}s,
E. Francisco, J. Andr\'{e}s, and J.\,M. Recio, Phys. Rev. B \textbf{64}, 184113 (2001).

\bibitem{lazzeri2} M. Lazzeri and A. Selloni, Phys. Rev. Lett. \textbf{87}, 266105 (2001).

\bibitem{Vittadini} A. Vittadini and A. Selloni, J. Chem. Phys. \textbf{117}, 353 (2002).

\end{thebibliography}
\end{document}